\documentclass[sn-apa,iicol]{sn-jnl}% APA Reference Style
\jyear{2022}%

\theoremstyle{thmstyleone}%
\usepackage{amsmath}
\usepackage{xcolor}

\theoremstyle{thmstyletwo}%

\theoremstyle{thmstylethree}%

\begin{document}

\title[ ]{  \textcolor{blue}{\hspace{4 cm}This is a pre-print}  \\  \\\textbf{On the modern deep learning approaches for precipitation downscaling}}


\author*[1]{\fnm{\normalsize Bipin} \sur{Kumar}}\email{bipink@tropmet.res.in}
\author[2]{\fnm{\normalsize Kaustubh} \sur{Atey}}
\author[1]{\fnm{\normalsize Bhupendra} \sur{Bahadur Singh}}
\author[1,3]{\fnm{\normalsize Rajib} \sur{Chattopadhyay}}    
\author[4]{\fnm{\normalsize Nachiket} \sur{Acharya}}    
\author[1]{\fnm{\normalsize Manmeet} \sur{Singh}}    
\author[5]{\fnm{\normalsize Ravi} \sur{S. Nanjundiah}}    
\author[1]{\fnm{\normalsize Suryachandra} \sur{A. Rao}}

\affil[1]{\orgname{\small{Indian Institute of Tropical Meteorology, Ministry of Earth Sciences, Government of India}}, \orgaddress{\small{Dr. Homi Bhabha Road, Pashan, Pune, India, 411008 \vspace*{0.1cm}}}}

\affil[2]{\orgname{\small Indian Institute of Science Education and Research}, \orgaddress{Dr. Homi Bhabha Road, Pashan, Pune, 411008, India \vspace*{0.1cm}}}

\affil[3]{\orgname{\small Indian Meteorological Department, Ministry of Earth Sciences, Government of India}, \orgaddress{Shivaji Nagar, Pune, 411005, India \vspace*{0.1cm}}}

\affil[4]{\orgname{\small CIRES, University of Colorado Boulder, and NOAA/Physical Sciences Laboratory}, \orgaddress{Boulder, Colorado, USA \vspace*{0.1cm}}}

\affil[5]{\orgdiv{\small Center for Atmospheric and Ocean Sciences}, \orgname{Indian Institute of Science}, \orgaddress{Bengaluru, India, 560012, India \vspace*{0.1cm}}}


\abstract{Deep Learning (DL) based downscaling has become a popular tool in earth sciences recently. Increasingly, different DL approaches are being adopted to downscale coarser precipitation data and generate more accurate and reliable estimates at local (~few km or even smaller) scales. Despite several studies adopting dynamical or statistical downscaling of precipitation, the accuracy is limited by the availability of ground truth. A key challenge to gauge the accuracy of such methods is to compare the downscaled data to point-scale observations which are often unavailable at such small scales. In this work, we carry out the DL-based downscaling to estimate the local precipitation data from the India Meteorological Department (IMD), which was created by approximating the value from station location to a grid point. To test the efficacy of different DL approaches, we apply four different methods of downscaling and evaluate their performance. The considered approaches are (i) Deep Statistical Downscaling (DeepSD), augmented Convolutional Long Short Term Memory (ConvLSTM), fully convolutional network (U-NET), and Super-Resolution Generative Adversarial Network (SR-GAN). A custom VGG network, used in the SR-GAN, is developed in this work using precipitation data.  The results indicate that SR-GAN is the best method for precipitation data downscaling. The downscaled data is validated with precipitation values at IMD station. This DL method offers a promising alternative to statistical downscaling.}




\maketitle

\section{Introduction}\label{Introduction}
There is increasing demand for reliable weather and climate information at local scales for better management and preparedness of billions across the globe. The observational data (especially ground-based observations) lack the spatial resolution required to project meteorological information on a micro-scale. Similarly, the outputs from global models (GCMs/ESMs etc.) are often coarse and hence are less helpful in understanding the local weather and climate and the impacts on specific sectors, e.g., agriculture, food, water, etc.  There is a requirement for developing more accurate rainfall datasets at finer scales especially over India. Although India Meteorological Department (IMD) has good coverage of stations over the country, the best available gridded data is still only at 0.25 x 0.25 km resolution. A long-pending question in this context is whether we can go for even higher resolution gridded data reconstruction using the available station observations. 

 To cater  such demands for high-resolution weather and climate data, forecasts, and projections are being produced using state-of-the-art, statistical and deep learning models \citep{Bano-Medina2020, Duben_Bauer2018, Kumar2021, Reichstein2019, Singh2021} . These datasets are produced through a combination of different methods, each having its own assumptions, advantages, and disadvantages. The methodology to obtain the high-resolution weather and climate information assumes that the local climate is a result of interactions between large-scale atmospheric characteristics (circulation, temperature, moisture, etc.) and local features (topography, water bodies, land use - land cover, etc.). One of the most popular techniques to derive weather and climate information at finer scales is done through downscaling. The downscaling is a technique for adding additional information to coarse datasets in order to obtain more trustworthy data at a finer scale, which is better at capturing regional and local inhomogeneities. 

 Although the information can be downscaled on both spatial and temporal dimensions, often it is the spatial information that is sought after in the weather and climate science and allied sectors. Regional models utilize the outputs from global models to produce dynamical downscaling by taking the large-scale atmospheric and oceanic conditions at the lateral boundaries and incorporating complex topography, the land-sea contrast, surface heterogeneities, and detailed descriptions of physical processes in order to generate realistic climate information at regional scales. A key limitation of this approach is its reliance on the accuracy of global model fields and inherent biases \citep{Seaby2013}. Thus, it requires a bias correction before going for further downscaling. Furthermore, the precipitation product is often quite erroneous in global as well as regional models. The observation data are generally downscaled using different interpolation techniques \citep{Pai2014,Rajeevan2006} . This issue can be addressed further using an alternative approach known as Statistical Downscaling (SD), which involves the establishment of an empirical relationship between historical/processed data (such as station data to the grid) and real-time/observational data (such as weather station data to the grid). Here, we explore the feasibility and reliability of generating high-resolution gridded data, prepared using ground observation,  based on deep learning techniques.

Deep Learning (DL) based downscaling is one type of SD that can be achieved using a variety of methods. \cite{Dong2015} proposed a method for image super-resolution. The same concept was later followed by   \cite{Vandal2017a} to produce high-resolution precipitation data. \cite{Kumar2021} used this method to downscale the gridded precipitation data on Indian landmass regions. For similar data using more input variables, \cite{Harilal2021}  applied augmented Conv LSTM.  Despite several studies adopting dynamical or statistical downscaling of precipitation, the accuracy is limited by the availability of ground truth. In this regard, a big question to ask is to check whether the downscaled grided data can be matched to point-scale observations. This is where the whole issue of reliable local precipitation information gets stuck and modern methods often fail. 

 We carry out the DL-based downscaling in this work to estimate the local projection of precipitation data from the India Meteorological Department (IMD), which was created by approximating the value from station location to grid point. We apply four different approaches to evaluate their performance and present the results of the analysis through robust statistical estimates. The considered approaches are (i) Deep Statistical Downscaling (DeepSD), augmented Convolutional Long Short Term Memory (ConvLSTM), fully convolutional network (U-NET), and Super-Resolution Generative Adversarial Network (SR-GAN).

The paper is organized as follows. The downscaling methods used in this work are detailed in the methodology section \ref{Methods} followed by the data pre-processing procedure in section \ref{Data_process_and_train}. Results are presented in section \ref{Results} and section \ref{Conclusion_and_Discussion} provides the conclusion and discussions on this work. 

\section{Methods}\label{Methods}

A statistical downscaling (SD) can be formulated as a functional mapping between low resolution (LR) data and corresponding high resolution (HR) data. A SD model takes $X$: LR  data as input and generate $\hat{Y}$: the high-resolution data as output

\begin{equation}
\label{Equation:SD_Formula}
    \hat{Y} = F(\Theta, X)
\end{equation}

The objective is to find a set of parameters $(\Theta)$ such the loss $\mathcal{L}(Y, \hat{Y}) $ between the generated and the ground truth data $(Y)$ is minimized. There are several methods available in the literature for statistical downscaling; a few, popular methods are discussed in this section. 

 We began with the DeepSD \citep{Vandal2017a}  algorithm, which is an enhanced version of the Super-Resolution Convolutional Neural Network (SRCNN) \citep{Dong2015} method that employs multivariable input for model training. \cite{Kumar2021} used precipitation and topography as inputs, which resulted in improved performance. They concluded that the method outperforms alternatives such as linear interpolation, SRCNN, and layered SRCNN. \cite{Harilal2021}  described a method for data downscaling using a recurrent convolutional LSTM. This technique captures both temporal and spatial relationships by combining fully linked LSTMs with convolutions. By capturing both spatio-temporal relationships and integrating additional auxiliary variables into the input, the ConvLSTM-SR architecture outperforms DeepSD.

 The work by \citep{serif2021} proposed a method called ‘DCN U-Net’ which demonstrated a downscaling architecture based on encoder-decoders for highly frequent meteorological data such as precipitation. This approach learns Spatio-temporal functional mapping using U-Net architecture. Bilinear interpolation is used to spatially downscale the input LR data and supplements it with the topographical channel. This input tensor is supplied to the U-encoder Net's section, and the time variable is inserted into the latent space shortly before the model's decoder portion. All these model architectures were trained with the objective to minimize Mean Squared Error (MSE), Root Mean Squared Error (RMSE) or L1 norm as a loss function. 

Generative Adversarial Network (GAN) is another approach\citep{Goodfellow2020} to generate a high-resolution image from a low-resolution input.  GAN-based algorithms may describe uncertainty using contemporary machine-learning techniques that are mostly independent of a specific statistical assumption \citep{Gagne2020}. These approaches have also been applied recently to the time evolution of atmospheric variables \citep{Scher2018}. There are several versions of GAN approaches, one of them is stochastic super-resolution GAN \citep{Leinonen2021} which provides an ensemble of plausible high-resolution outputs for a given input. 

 Another popular GAN-based algorithm, proposed by \citep{Ledig2017}, has been shown to be more successful in generating the super-resolution of a single picture than the previous one. This algorithm is called SRGAN (Super-Resolution Using Generative Adversarial Network) and it is used for single picture super-resolution tasks. It is built on a GAN and uses a deep learning architecture to obtain its results. However, even when dealing with extremely large scaling factors, it is still possible to retrieve the finer elements of the original image. The SRGAN method has been employed in this work, and its performance compared to that of three other algorithms (DeepSD, ConvLSTM and U-NET) used for ML-based data downscaling.

 Table \ref{tab:training_details} shows the architecture of all four models. These models were trained on the LR ($1^o$) IMD data for $4x$ resolution enhancement and we compared the results using correlation as a metric.  Deep learning algorithms for data downscaling are frequently used in conjunction with a loss function based on the Mean Squared Error (MSE) \citep{Kumar2021, Harilal2021}. Due to the point-wise average of the solutions, one of the biggest disadvantages of utilising an MSE-based loss function is that the resulting solution seems excessively smooth. SRGAN solves this problem by employing a unique loss function known as perceptual loss, which is useful in adversarial training situations. Also, it  helps to recover the final details of the data. When calculating the perceptual loss, Ledig et al. (2017) use a weighted sum of content and adversarial loss content as shown in equation \ref{Equation:Perceptual_Loss} below

\begin{equation}
\centering
{\label{Equation:Perceptual_Loss}
l^{SR} = \underbrace{l_{X}^{SR}}_{content \hspace{3pt} loss} + \underbrace{10^{-3} l_{Gen}^{SR}}_{adversarial \hspace{3pt} loss}}
\end{equation}

\begin{table*}[ht]
   \centering   
 \caption{List of hyper parameters used in the four models.  In all models, the Adaptive Moment Estimation (ADAM) optimizer was used which provided faster convergence.}
\begin{tabular}{|p{1.55cm}|p{4.9cm}|p{1.2cm}|p{1.7cm}|p{0.8cm}|}
     \hline  
    \textbf{Model} & \textbf{Activation Function} & \textbf{Learning Rate} & \textbf{Loss \newline Function} & \textbf{Batch Size}\\ \hline
    
     DeepSD & PRELU (Parametric Rectified Linear Unit) & 0.001 & MSE & 200 \\ \hline
    
     Augmented ConvLSTM & Conv-2D- RELU & 0.0003 & RMSE & 8 \\ \hline
    
     UNet & Conv-3D- RELU & 0.001 & Grad Loss \newline (Custom) & 8 \\ \hline   
    
     SRGAN & {}Generator- Residual Blocks – PRELU
\newline Generator- UpSampling Blocks- PRELU
\newline Discriminator- Conv Body- Leaky RELU
\newline Discriminator- Dense 1-  Leaky RELU
\newline Discriminator- Dense 2-  Sigmoid
 & 0.0001 & Perceptual Loss \newline (Custom) & 256 \\ \hline   
\end{tabular}
    \label{tab:training_details}
\end{table*}

\subsection{The Kriging method}\label{Kriging_Method}

In this study, the downscaled data obtained from DL method has been verified from actual station data by approximating the precipitation values at grid point to nearest station. Kriging method \citep{Oliver1990} was used to interpolate the precipitation values at grid points to station locations. In the kriging approach, statistical correlations among the observed points are used to generate the values. For a given grid point $u_{i, j}$ a function $f$  to approximate the value from grid point to station location can be written as

\begin{equation}
\label{Equation:Kriging_Eq1}
f(x, y, z) = \frac{\sum^{i=0}_{n} w_i u_i}{\sum^{i=1}_{n} w_i}
\end{equation}

\noindent Where $(x, y)$ are latitude and longitude and $z$ represents the elevation of a particular grid point, $n$ is the total number of grid points in a specified radius (see figure \ref{fig:clm_analysis}). The weights $w_i$ are determined in terms of the distance $d$ between the station location and the grid point. Now consider a vector $u_0 = uw^T$, $w$ is the weight vector calculated as $Aw=b$, where,

\begin{equation}
	\label{Equation:Kriging_Eq2}
	\begin{split}
A = r(x_i,x_j ) \hspace{0.5cm} b = r(x_0,x_i) \hspace{0.5cm}  \text{and}  \\
\hspace{0.5cm} r = 1/2 (u(x+h)-u(x))^2	
\end{split}
\end{equation}

\noindent Here, $x_0$ represents location of point of interest and the respective precipitation value at that point is $u_0$. Then the approximation function (equation \ref{Equation:Kriging_Eq1}) can be written as 

\begin{equation}
    \label{Equation:Kriging_Eq3}
\hat{f} = u w^T
\end{equation}

The grid values are approximated using equation \ref{Equation:Kriging_Eq3}.

\section{Data Pre-processing and Training}\label{Data_process_and_train}

The IMD gridded data \citep{Pai2014,  Rajeevan2008} \cite{Rajeevan2008} was used for training in this study. This data is available in two resolutions: ($1^ox1^o \sim $ $110 \,km \times 110 /, km$) and ($0.25^o \times 0.25^o$ $\sim$ $25 \, km \times 25 \, km $). The $1^o$ data has a dimension of (33,35) and 0.25 of (129,135). We have mainly used the $0.25^o$ data which was downsampled to $1^o$ data. 
 This process was carried out because of the non-similarity in both data. In particular, the low-resolution data was prepared using fewer ground stations available at that time. Gradually, the number of stations in 2014 increased, and the high-resolution data was prepared using the greater number of ground stations. The downsampling was carried out using following two steps:

\begin{enumerate}
    \item Average pooling with (2,2) kernel
    \item Max pooling with (2,2) kernel
\end{enumerate}

The minimum and maximum precipitation values are calculated by grouping the same days for all the years and min-max normalization is employed to scale data in the range of [0,1]. The formula used for normalization is 
\begin{equation*}
Z = \frac{X - X_{min}}{X_{max} - X_{min}}
\end{equation*}

\noindent The details of the time period for the data is provided in table \ref{tab:training_details}. 

\noindent In addition to the IMD gridded data, we have also taken into consideration the IMD station data over the period of time (2005-2009) in order to validate the DL model output. The specifics of the stations from which the data was gathered are presented in section \ref{results_subsection_3}. %4.2.

\begin{table}[ht]
   \renewcommand{\arraystretch}{1.5}
    \centering   
    \caption{Details of the IMD gridded data used for training and testing (also referred as ground truth (GT) data)}    
    \begin{tabular}{|c|c|c|} \hline    
    \textbf{Data} & \textbf{Training} & \textbf{Testing}\\ \hline    
     IMD & 1975 - 2004 & 2005 - 2009  \\ \hline
\end{tabular}
    \label{tab:data_details}
\end{table}

\subsection{VGG training for precipitation data}\label{VGG_Training}

As mentioned in the section \ref{Methods}, the SR-GAN method uses a unique loss function, called perceptual loss which has two components: content loss and adversarial loss.  The content loss (equation \ref{Equation:Content_Loss}), also known as the VGG (Visual Geometry Group) Loss, is defined as the Euclidean distance between the feature representations of the produced data $\hat{Y}$ and the corresponding ground truth data $(Y)$. The VGG network (Simonyan and Zisserman, 2015) was developed by a group at the University of Oxford (https://www.robots.ox.ac.uk/~vgg/). The data is fed into a pre-trained VGG network, which produces feature representations. The content/VGG loss is defined as (Ledig et al., 2017)

\begin{equation}
    \label{Equation:Content_Loss}
    \begin{split}    
    l_{VGG/i,j}^{SR} = \frac{1}{W_{i,j} H_{i,j}} \sum_{x=1}^{W_{i,j}} \sum_{y=1}^{H_{i,j}} \\ (\phi_{i,j}(Y)_{x,y} - \phi_{i,j}(\hat{Y})_{x,y})^2 
    \end{split}
\end{equation}

\begin{figure*}[]
	\centering
	\includegraphics[scale=0.09]{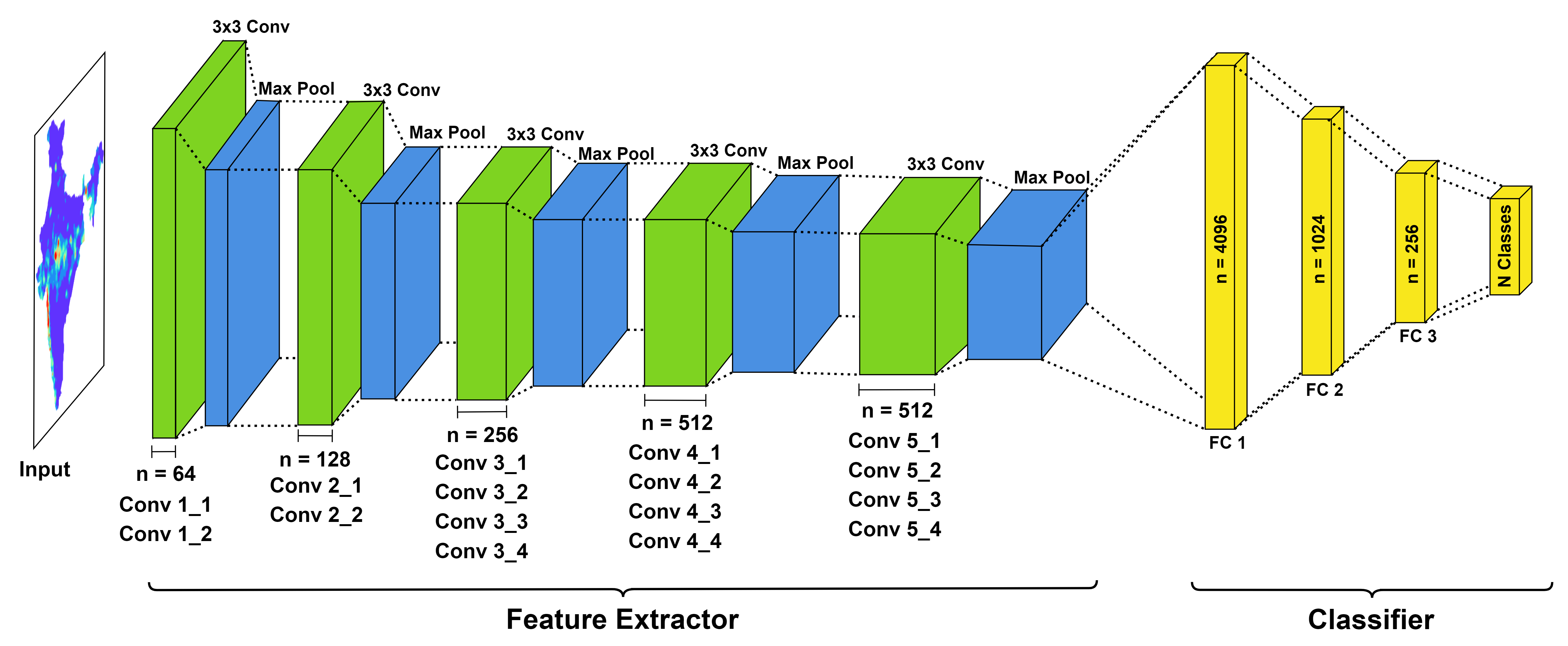}
	\caption{The details of the VGG training architecture used for precipitation data.}
	\label{fig:VGG}
\end{figure*}

Here $W_{i,j}$ and $H_{i,j}$ are the dimension of the respective feature map within VGG network The $\phi_{i, j}$ are the feature extraction map and $Y$ and $\hat{Y}$ are target and predicted data (generated from SR-GAN) respectively.  Because they were trained on the ImageNet dataset, which consists of natural RGB images, we were unable to use the publicly available pretrained VGG networks in this study. Instead, we trained a VGG network on meteorological data in such a way that it learns a rich feature representation for meteorological variables through experimentation.  Thus, we developed a novel VGG model which can work on meteorological data.  A sketch illustrating the training architecture is provided in figure \ref{fig:VGG}. During SR-GAN adversarial training, this trained model computes the content loss component of perceptual loss. We used the same data for the VGG training which was used for model training, i.e. for the duration 1975-2004. It contains many layers to extract finer scale feature as depicted in figure \ref{fig:VGG}, more details can be found in \citep{simonyan2015}.

\subsection{Other Aspects of Training}\label{Other_Training}

The SR-GAN method contains two networks namely, generator and discriminator networks. The generator network is pretrained with an MSE loss function to get it initialized. During adversarial training, this helps the generator to avoid undesirable local optima. Finally, in an adversarial setting, the pretrained generator is trained alongside the discriminator network. The discriminator learns to differentiate between the generated output (SR) and the ground truth (GT). The VGG feedback and the discriminator feedback are used to calculate the perceptual loss. Backpropagation of the perceptual loss to the generator allows it to generate finer texture details. The details of all three training are provided in the table 3.

The SR-GAN model integrated with the newly developed pre-trained VGG model has been used in this work for the local projections of the IMD gridded data.

\section{Results}\label{Results}

\begin{figure*}[t]
	\centering
	\includegraphics[scale=0.4]{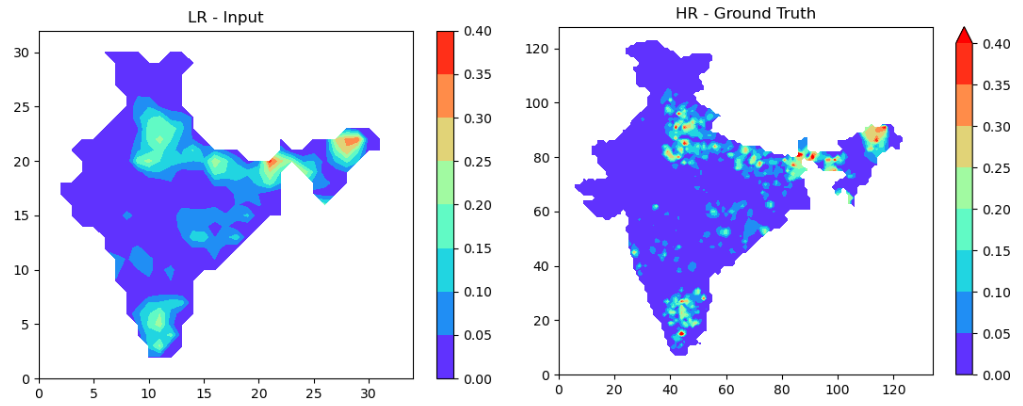}
	\caption{A 2D representation of 1° (low resolution) input data (Left) ; 0.25° (high resolution) target data (Right).}
	\label{fig:Data_Vis}
\end{figure*}

The data we utilised is IMD gridded data, as stated in the previous section. We applied four DL-based downscaling algorithms to this data, which are described in section \ref{Methods}. The provided data has been pre-processed using the methods listed above. Figure \ref{fig:Data_Vis} depicts a representative snapshot of the data in both low and high resolution.

\begin{figure}[t]
    \centering
    \includegraphics[scale=0.45]{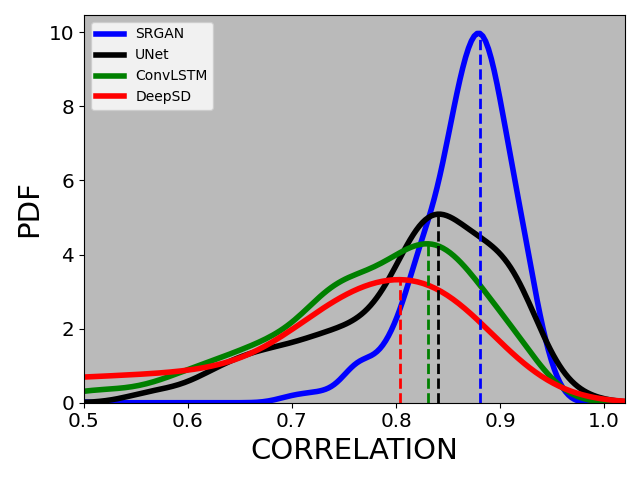}
    \caption{Comparison of correlation obtained using HR data and resulting data from the four methods for the time duration 2004-2009.}
    \label{fig:All_Correlation}
\end{figure}

\noindent The resulting downscaled data ($1^o$ to $0.25^o$) was compared to available HR ($0.25^o$) data. An illustration of the comparison of correlation PDFs obtained using these four approaches is shown in figure \ref{fig:All_Correlation}. As can be seen, the SR-GAN appeared to be the most effective approach in correlation coefficient comparison. Our custom pre-trained VGG feature extractor led the SR-GAN model to generate and recover higher frequency details. The correlation PDFs comparison confirms that SR-GAN is one of the best methods for DL downscaling among four methods discussed in this study. In the next step, we applied this method to generate high resolution (upto 4x resolution) gridded data.

\subsection{Generation of high-resolution data using IMD 0.25° data }\label{results_subsection_1}

\begin{figure*}[t]
	\centering
	\includegraphics[scale=0.07]{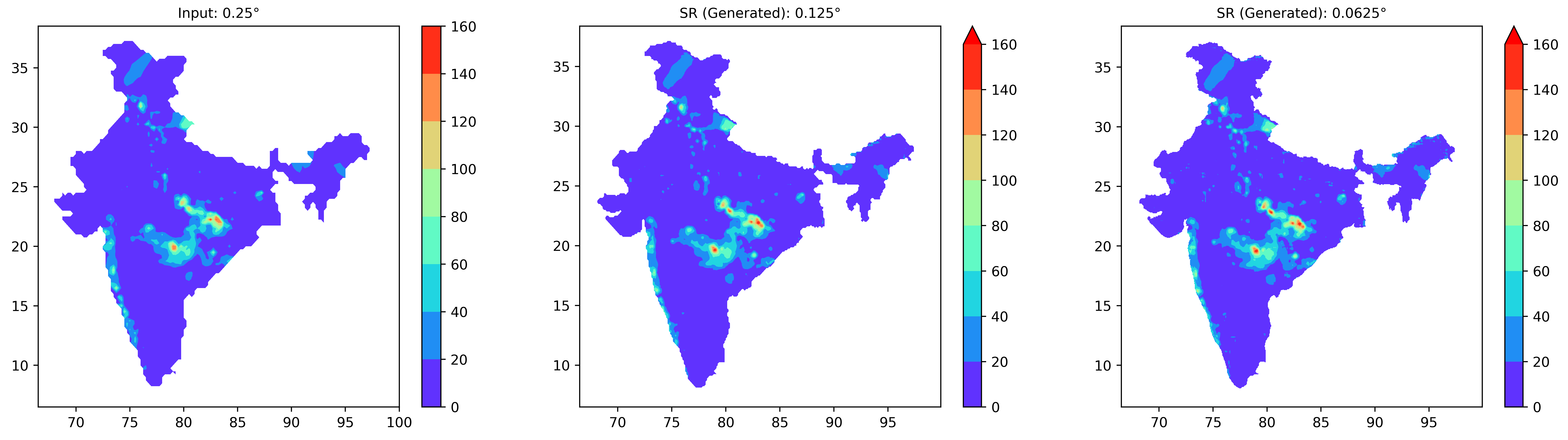}
	\caption{Visual representation of rainfall data downscaling up to 4x resolution for the day 05-08-2005.}
	\label{fig:Results_2x_4x}
\end{figure*}

We have determined that the SR-GAN method is the best among four methods, namely, DeepSD, ConvLSTM, U-NET and the SR-GAN. This method has been used to generate the high resolution IMD data, in particular from $0.25^o$, as input, to $0.125^o$ (2x) and $0.0625^o$ (4x) have been generated.

 The same (SR-GAN) model architecture has been used for the creation of this high-resolution data. The dataset's duration also remains same. Figure \ref{fig:Results_2x_4x} shows a visual comparison of rainfall data for a certain day, i.e., 5th August, 2005. The outputted datasets show localized rainfall patches in both the high resolutions data sets. We noted that these rainfall values were not present in the available IMD gridded data of 0.25°x0.25° resolution. In later section \ref{results_subsection_2}, we would check the accuracy of these rainfall patches with observed station rainfall values.

\subsection{Validating the high resolution output with station data}\label{results_subsection_2}

\begin{figure*}[ht]
    \centering
    \includegraphics[scale=0.62]{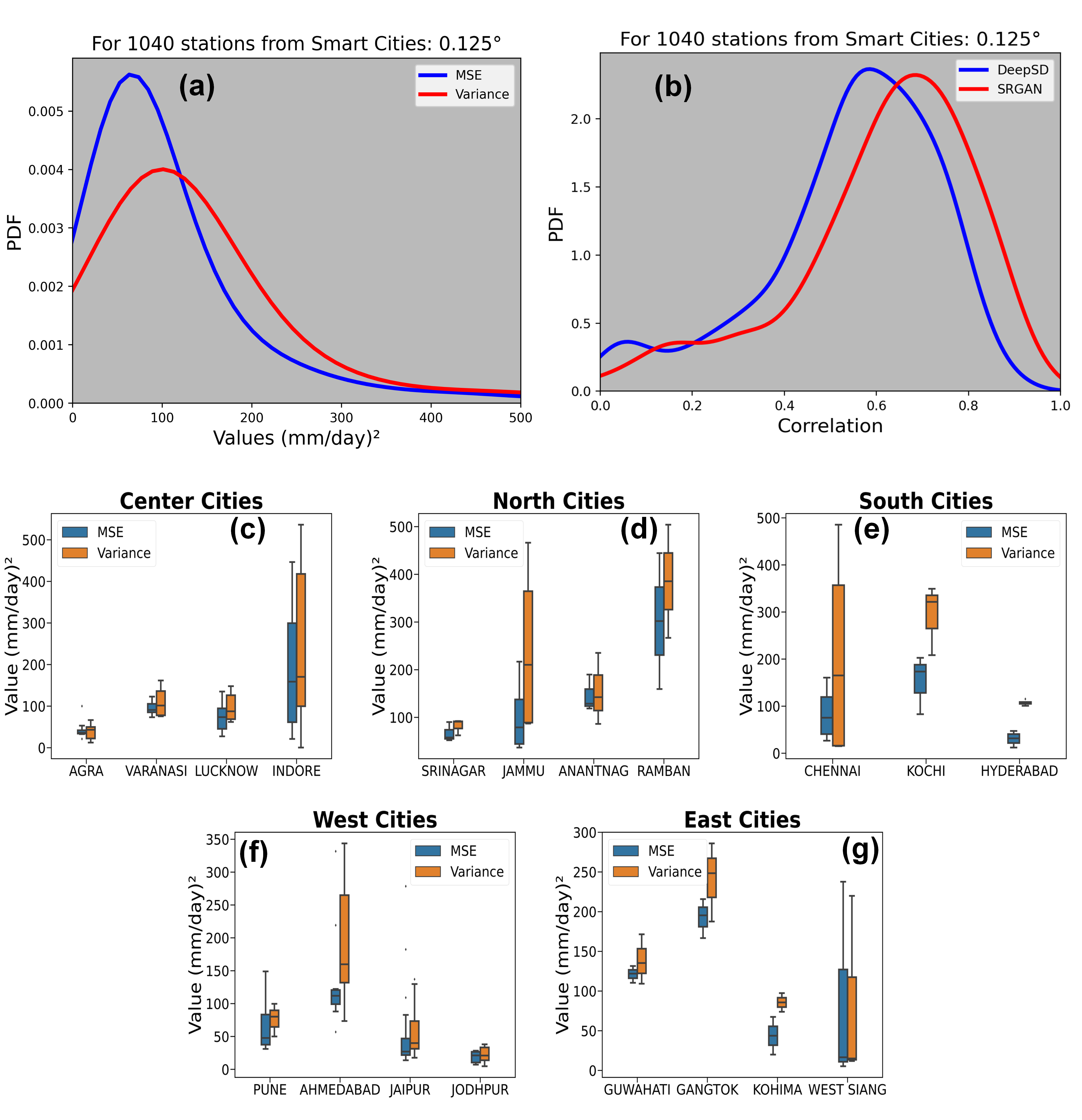}
    \caption{Comparison of MSE and variance (panel a) for 1040 stations from 78 smart cities of India for the $2x$ high resolution, i.e., $0.125^o$ data. Panel (b) depicts the correlation comparison obtained from downscaled gridded data the station data of the same 1040 stations using SRGAN and liner interpolation.  A comparison of variance and MSE for selected cities are presented in the panels (c) – (g).}
    \label{fig:Results_12km}
\end{figure*}

The high-resolution IMD gridded data ($0.125^o$ and $0.0625^o$) obtained using the SR-GAN method (see figure \ref{fig:Results_2x_4x}) has been validated with rainfall values at the ground station. The validations for both resolutions were performed by interpolating the downscaled data, using the Kriging method explained in section \ref{Methods}, to the station locations as well as comparing the MSE and correlation. We considered a few IMD ground observation stations located in the various smart cities of India \href{(https://smartcities.gov.in/)}{https://smartcities.gov.in/}. Although the government of India has named 100 cities as smart cities, IMD stations are present at only in 78 cities. The total 1040 IMD stations were found in these 78 smart cities. We used the actual precipitation values at stations from these cities to validate our high resolution gridded data obtained from AI model output. First, a  PDF of MSE (between station data and interpolated downscaled data of these  ground stations) for $0.125^o$ resolution data has been calculated and compared with variance in precipitation data  for the duration (2004-2009). This comparison is presented in figure \ref{fig:Results_12km} (a). Further, a correlation PDF comparison obtained by SR-GAN method and DeepSD, using the same data, is depicted in figure \ref{fig:Results_12km} (b).  It shows that the SR-GAN method is better in capturing the rainfall values. 

\begin{figure*}[ht]
    \centering
    \includegraphics[scale=0.62]{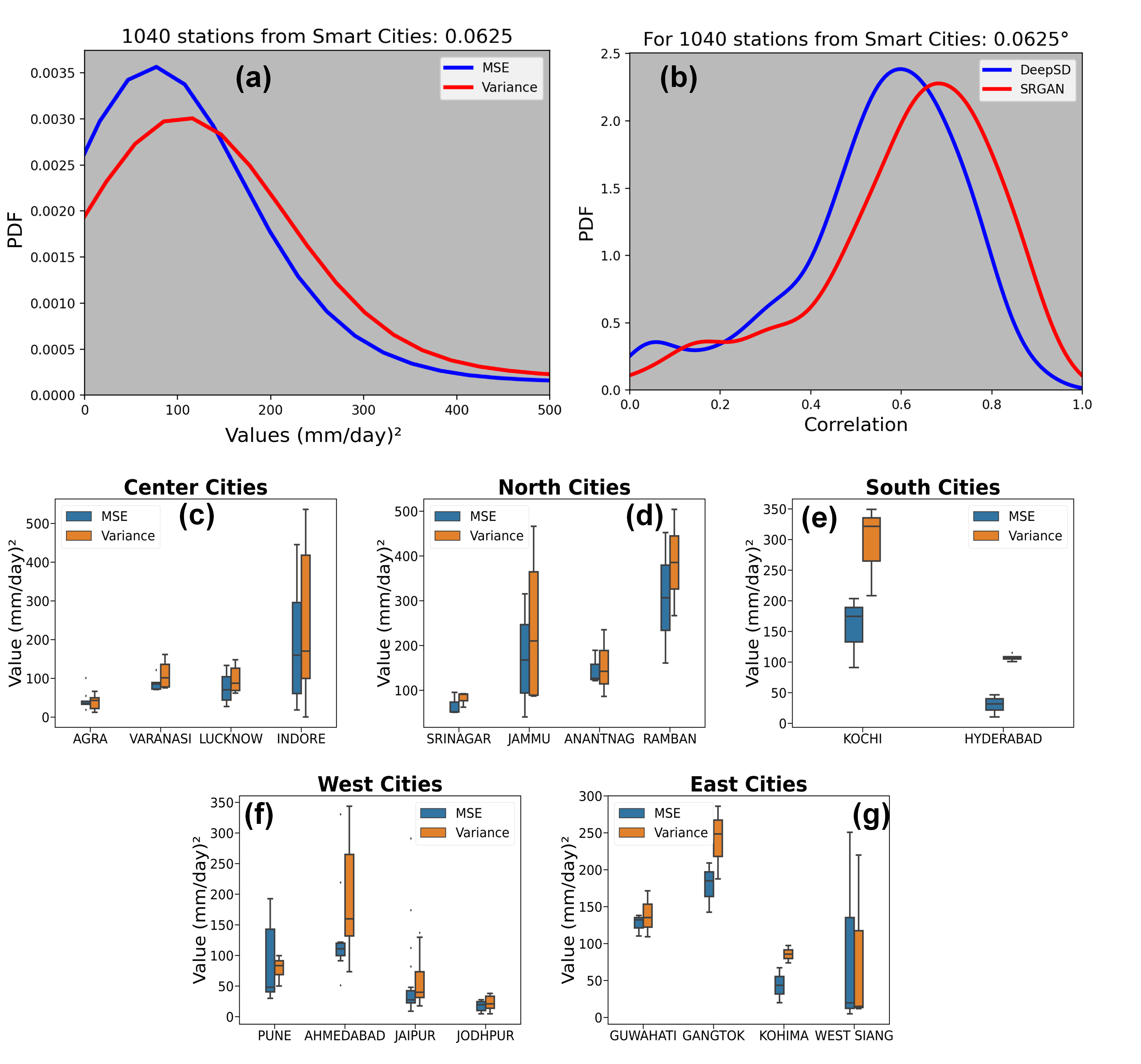}
    \caption{Similar comparison as in Figure \ref{fig:Results_6km} but for 4x resolution data (0.0625°).}
    \label{fig:Results_6km}
\end{figure*}

 Apart from the comparison of MSE and correlation PDF, we have validated the data for a few selected cities by comparing the MSE and variances in the data.  Here, the city selection was done based on data availability and spread across different geographic locations of the country. For robust quantifications, we chose the cities in such a way that each part has at least three stations or more. After obtaining the rainfall values on station location we calculated the MSE compared this with variance in the data for a number of cities in central India, as depicted in Figure \ref{fig:Results_12km} (c). A total of four cities were selected for the comparison in this panel, and in each case, the MSE is found to be smaller than the variation. In a similar manner, comparisons have been made for a number of other Indian towns located in the northern, southern, western, and eastern regions. As can be observed in figures \ref{fig:Results_12km} (d)- \ref{fig:Results_12km} (g), the MSE has been consistently lower than the variance in virtually all situations. This analysis validates correctness of the high-resolution data obtained from SRGAN. After validating the 2x resolution data with station values, we generated the 4x resolution data, i.e., for $0.0625^o$ and did the similar analysis. In this data, again we found the MSE value lower than the data variance. The values of MSE PDF compared with variance PDF for the same stations from the same number of smart Indian cities are presented in the figure \ref{fig:Results_6km}. 

 The analysis presented in figures \ref{fig:Results_12km} and \ref{fig:Results_6km} proved that the data obtained from the SR-GAN method is valid. The next step remains for us to validate the extra rainfall values appeared at the new locations (which didn’t exist in the original coarser grid) in the high-resolution map. The values at these locations have been compared with nearest IMD station. This process is discussed in the next section.

\subsection{Validating the additional information from ground stations }\label{results_subsection_3}

\begin{figure*}[htpb]
    \centering
    \includegraphics[scale=0.45]{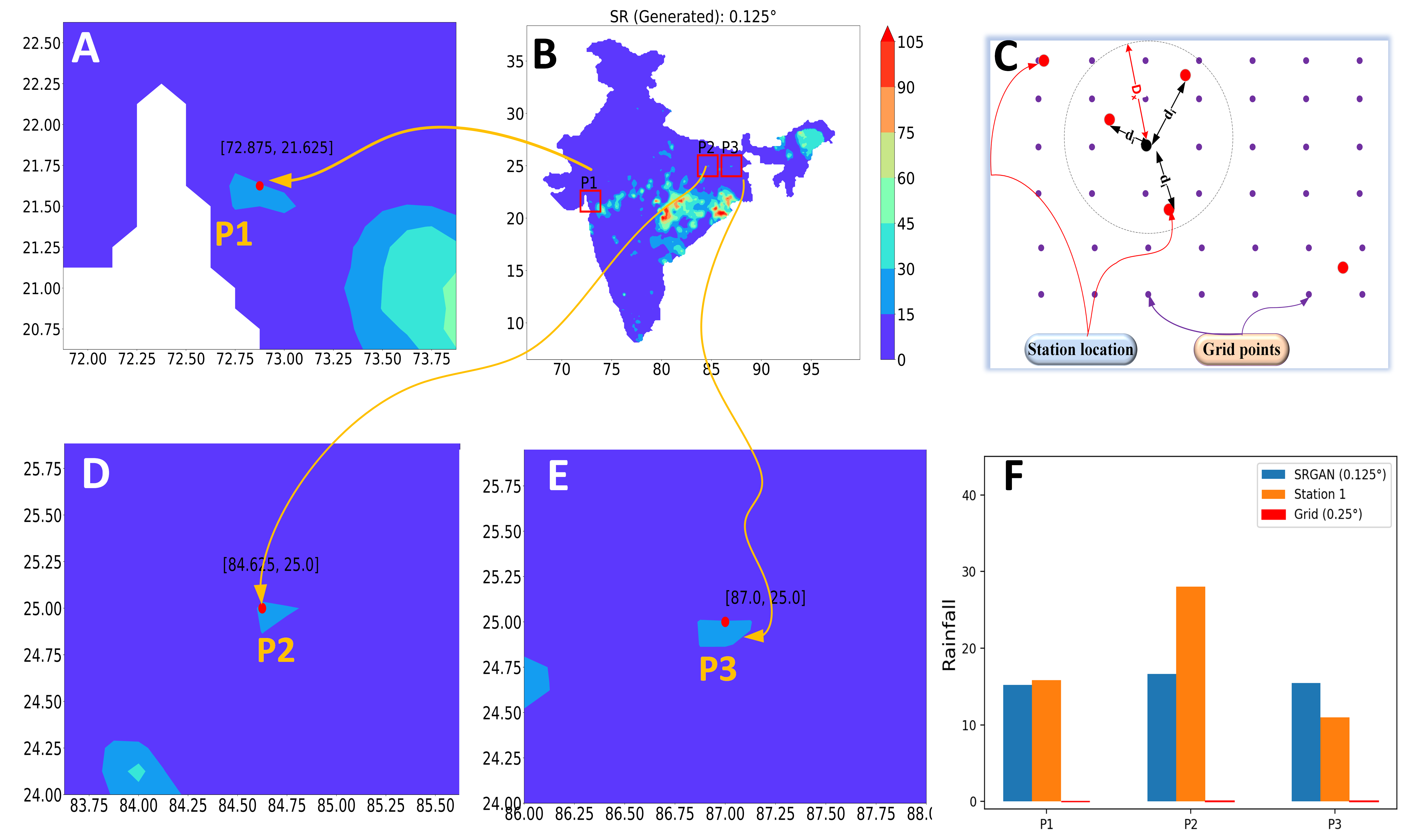}
    \caption{Validating the extra points obtained from the SRGAN method for high-resolution data.  The extra rainfall points obtained are shown in panel (B). The zoomed versions of those points are represented in panels (A, D, and E). We used a search radius of 12 km to locate the nearest ground station as depicted in panel (c). In this panel the blue dots represent grid points in a gridded data and the IMD station locations are represented by red dots.  Panel (F) represents a comparison of the predicted rainfall and values at the nearest stations.}
    \label{fig:extra_points}
\end{figure*}

We did this comparison exercise so as to determine whether the model generated ‘false positives’ or in fact it is reproducing the real data. The 2D map of the high-resolution ( $0.125^o$) data as depicted in the \ref{fig:Results_2x_4x} has additional grids points representing the rainfall values which are absent in the original data (see the first panel). To verify the legitimacy of additional rainfall appearing on the finer grid, we collected the coordinates of those points, located the nearby stations, and compared the rainfall of those stations. Panel (B) of figure \ref{fig:extra_points} indicates some of the additional rainfall patches obtained from the high-resolution data. Their zoomed locations (with Lat, Lon) are shown in panels (A, D, and E). These extra patches were validated with the actual rainfall values at nearby IMD ground stations.

First, the stations lying in the area of 12 km radius were located then the precipitation values at grid points were approximated to station location using Kriging method described in section \ref{Kriging_Method}.  The approximated values were compared to the actual observation rainfall values at that station. The panel (f) of figure \ref{fig:extra_points} depicted such a comparison for precipitation values on a particular day 05-08-2005. It was found that the precipitation was recorded to those IMD stations on a particular day which was not present in the original gridded data of  $0.25^o$ resolution. Thus, the downscaling method was perfectly validated with the actual observation.  

\subsection{Climatology Analysis}\label{clm_results_subsection_4}

\begin{figure*}[htpb]
    \centering
    \includegraphics[scale=0.45]{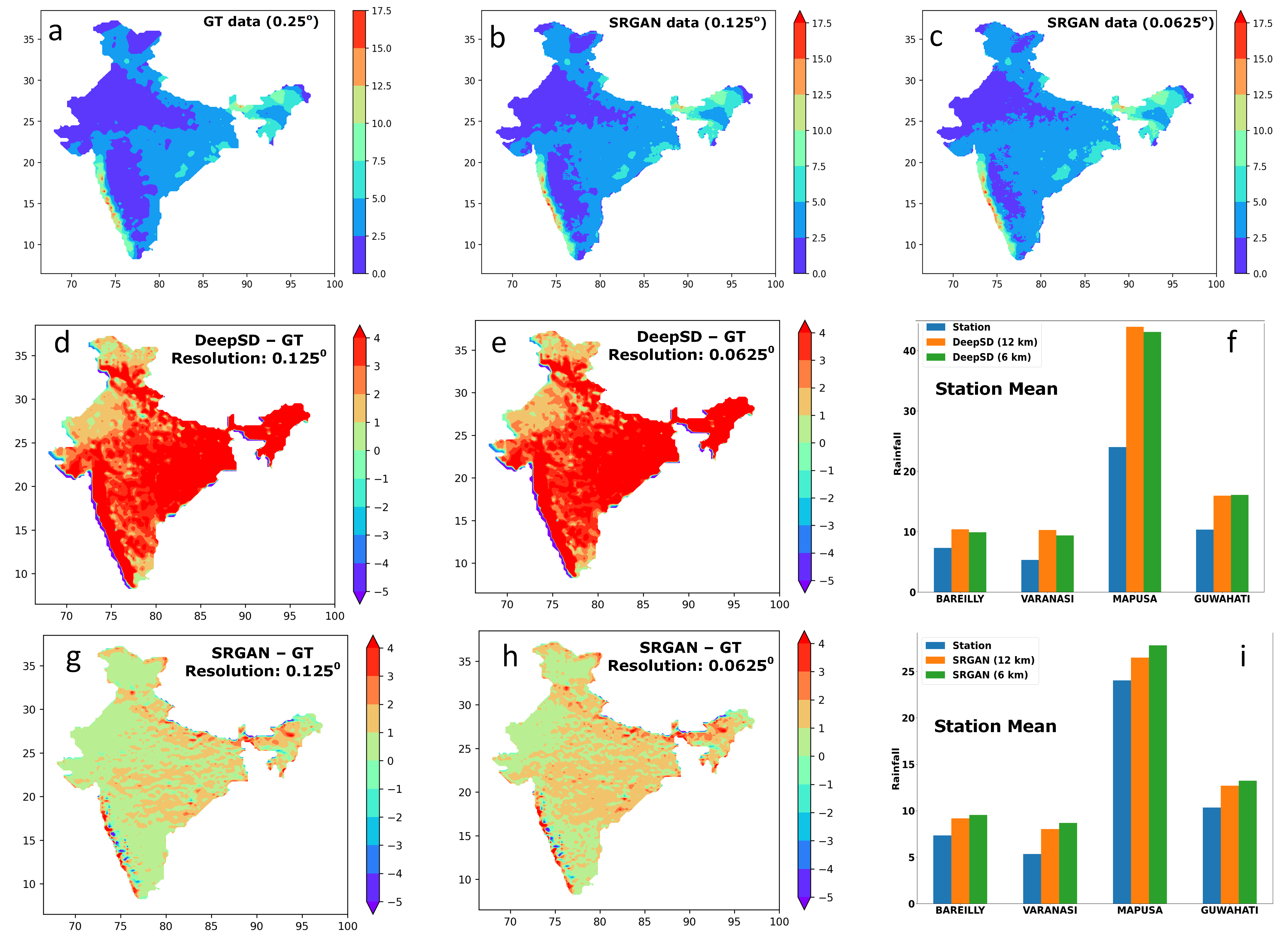}
    \caption{The climatology analysis of downscaled data. Panels (a)-(c) depict the climatology for test data (2005-2009) i.e. IMD gridded data  (indicated as GT), SR-GAN ( $0.125^o$) and SR-GAN ( $0.0625^o$). Panels (d) and (e) represent climatological biases for $2x$ and $4x$ resolution data for DeepSD and the panel (f) compares the station mean values with DeepSD generated values at those locations. Similarly, panels (g) and (h) have SR-GAN climatological   biases and (i) is for comparison with SR-GAN generated mean values.  These biases are calculated using original 0.25° data.}
    \label{fig:clm_analysis}
\end{figure*}

Using the input data, 2x resolution and 4x resolution data (obtained from the SR-GAN method), we calculated the climatology for the test data time duration (i.e., 2005-2009) and compared with it with DeepSD model as reference. The spatial plots for all three climatology datasets are depicted in figure \ref{fig:clm_analysis} (panels a-c). They show that the small-scale spatial variations are visible in high resolution data, while larger patterns are preserved in downscaled data. Furthermore, we calculated the biases in climatology using the original input data and DeepSD and SR-GAN generated data. The bias comparison in panels (d) and (e) show that the DeepSD model is overestimating the climatology at almost all the locations. Even the actual precipitation values at stations location are overestimated as depicted in the panel (f). The panels (g) and (h) bespeaks that the climatology in the Western Ghats and North-East area is enhancing using SR-GAN generated high resolution data. Moreover, the precipitation values at station locations are quite close to the actual observation values (see the panel (i)). We note that DeepSD generally overestimates the rainfall by about 25\% whereas the SR-GAN estimates are closer, and the bias is further reduced by half. Thus, the SR-GAN is the best method for precipitation downscaling as evident through climatological bias analysis.

\section{Conclusion and Discussion}\label{Conclusion_and_Discussion}

Downscaling of low-resolution data has several practical applications. A variety of statistical downscaling methods are available to generate high-resolution data from low-resolution gridded data.  In this study, we used the deep learning method to downscale the gridded precipitation data from IMD. We used low-resolution gridded data ($\sim$ $25 \, km x 25 \,km$) as input to make $2x$ ($\sim$ $12 \,km \times 12 \,km$) and 4x ($\sim$ $6 \, km \times 6 \,km$) high-resolution data. Four different types of deep learning methods were used to downscale precipitation data. Previously published research by \cite{Kumar2021} and \cite{Harilal2021} employed two different methods to accomplish this objective, and we followed their lead. By the comparison of correlation coefficients between these two investigations, \cite{Harilal2021} concluded that the augmented ConvLSTM approach outperformed the DeepSD method which was employed by \cite{Kumar2021}. In this work, we built on the findings of the previous two studies and utilised two more approaches, namely U-NET and SR-GAN. Due to the point-wise average of the solutions, one of the biggest disadvantages of utilising an MSE-based loss function in DeepSD and ConvLSTM methods is that the resulting solution seems excessively smooth. SRGAN based approach solves this problem by employing a unique loss function which perform better in capturing the spatial features. 

 It was discovered that the initial VGG model, used in the SR-GAN algorithm, given by the developer was trained on an image collection and did not perform well on the precipitation data. In particular, the correlation value was quite poor  as well as the fine details of precipitation values were not captured. Therefore, we constructed a unique VGG model, which we trained using precipitation data to achieve our goals. Particular attention was paid to the correlation coefficient metric, and it was shown that SR-GAN along with the developed custom VGG model is the most data-driven technique for statistical downscaling of precipitation data. Based on the correlation coefficient comparison, we concluded that the SR-GAN was the best DL model among the four models utilized and generated data at a resolution of two and four times. In addition to this, we have shown some specific cases in which the SR-GAN method can be used to enhance the rainfall amplitude (Fig. \ref{fig:extra_points}). It looks like the enhancement in station rainfall amplitude through SR-GAN reduces the overall MSE (Fig. \ref{fig:Results_12km} and \ref{fig:Results_6km}). 

 When comparing 2x resolution data to low-resolution data, the 2x resolution data contained additional points (refer to Figures \ref{fig:Results_2x_4x} and \ref{fig:extra_points}). We considered the precipitation values from a few nearby ground stations and compared them to the rainfall values at those stations in order to validate these additional points. To find the ground station locations, we searched an area with a radius equal to or less than the grid size. For example, in the instance of 0.125° data (approximate 12.5 km grid resolution), we searched an area with a radius of 12 km to find the ground station locations as shown in figure \ref{fig:extra_points}.  Using the high-resolution data, we found that there are precipitation values available at the locations of additional points that were not present in the actual low-resolution data. As a result, the high-resolution data has been validated using station values. We attempted to create data with a 4x resolution and obtained additional spots on the map. This has also been verified with the help of station data. We note that the MSE of the high-resolution data is much lower than the data variance. The MSE further reduces when we go to even higher resolution data.

\section*{Declarations}

\textbf{Conflicts of interest}: The authors have no conflicts of interest in this work.  \\

\noindent \textbf{Author Contribution}: BK, BBS, and RC have conceptualized the idea and contributed to the manuscript preparation. KA develop the code for the DL methods. RSN contributed to preparing the manuscript.  NA, MS and SA Rao helped with manuscript writing. \\

\noindent \textbf{Data Availability}: The gridded data, used in this study, can be obtained from IMD website \href{(imdpune.gov.in)}{imdpune.gov.in}. Any other data may be made available from the corresponding author on a reasonable request. \\

\noindent \textbf{Funding statement}: No funding was received for this study. \\

\noindent \textbf{Acknowledgment}: IITM Pune is funded by the Ministry of Earth Science (MoES), the Government of India. This work was done using HPC facilities provided by MoES at IITM Pune.


\begin{footnotesize}
\bibliography{downscale_ref}% common bib file
\end{footnotesize}



\end{document}